%% file: 00-omnifill.tex
\documentclass[sigconf]{acmart} 
\newtoggle{manuscript_version}
 \togglefalse{manuscript_version}

\AtBeginDocument{%
  }

\usepackage{tabularx}

\acmYear{2023}
\setcopyright{none} 
\acmDOI{}
\acmISBN{}
\acmConference[arXiv Preprint]{arXiv Preprint}{October 2023}{}
\acmBooktitle{arXiv Preprint}

\begin{document}
\title{OmniFill: Domain-Agnostic Form Filling Suggestions Using Multi-Faceted Context}

\author{Timothy J. Aveni}
\email{tja@berkeley.edu}
\orcid{0000-0002-6549-5249}
\affiliation{%
  \institution{University of California, Berkeley}
  \city{Berkeley}
  \state{California}
  \country{USA}
  \postcode{94720}
}

\author{Armando Fox}
\email{fox@berkeley.edu}
\orcid{0000-0002-6096-4931}
\affiliation{%
  \institution{University of California, Berkeley}
  \city{Berkeley}
  \state{California}
  \country{USA}
  \postcode{94720}
}

\author{Björn Hartmann}
\email{bjoern@berkeley.edu}
\orcid{0000-0002-0693-0829}
\affiliation{%
  \institution{University of California, Berkeley}
  \city{Berkeley}
  \state{California}
  \country{USA}
  \postcode{94720}
}

\renewcommand{\shortauthors}{Aveni et al.}

\begin{abstract}
    Predictive suggestion systems offer contextually-relevant text entry completions.
	Existing approaches, like autofill, often excel in narrowly-defined domains but fail to generalize to arbitrary workflows.
	We introduce a conceptual framework to analyze the compound demands of a particular suggestion context, yielding unique opportunities for large language models (LLMs) to infer suggestions for a wide range of domain-agnostic form-filling tasks that were out of reach with prior approaches.
	We explore these opportunities in OmniFill, a prototype that collects multi-faceted context including browsing and text entry activity to construct an LLM prompt that offers suggestions \textit{in situ} for arbitrary structured text entry interfaces.
	Through a user study with 18 participants, we found that OmniFill offered valuable suggestions and we identified four themes that characterize users' behavior and attitudes: an “opportunistic scrapbooking” approach; a trust placed in the system; value in partial success; and a need for visibility into prompt context.
\end{abstract}

\ccsdesc[500]{Human-centered computing~User interface design}
\ccsdesc[500]{Human-centered computing~Web-based interaction}
\ccsdesc[500]{Human-centered computing~Natural language interfaces}
\ccsdesc[500]{Human-centered computing~Text input}

\keywords{large language models, intelligent user interfaces, form filling, context-awareness, Web automation}

\begin{teaserfigure}
  \includegraphics[width=\textwidth]{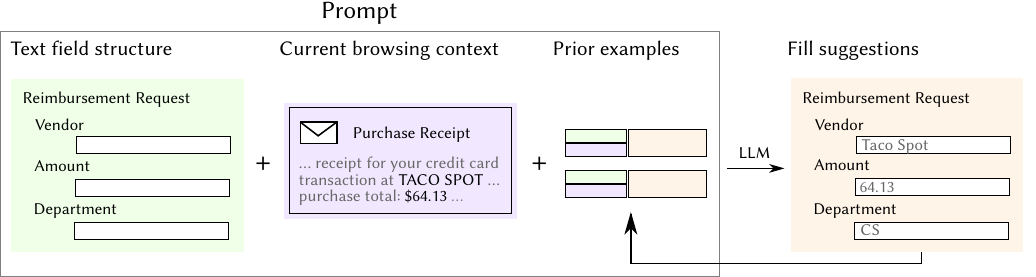}
  \caption{OmniFill offers domain-agnostic suggestions for Web forms using a multi-faceted prompt to a large language model.}
  \Description{A box showing a ``Reimbursement Request'' form (labeled ``Text field structure'', a box showing abbreviated contents of a purchase receipt email (labeled ``Current browsing context'', and a third box (labeled ``Prior examples'') are all outlined with a box labeled ``Prompt''. An arrow from the prompt to a fourth box is labeled ``LLM''. The fourth box is labeled ``Fill suggestions'' and contains the form from the first box, but filled with completion suggestions. The final box has an arrow drawn in the direction of the ``Prior examples'' box.}
  \label{fig:teaser}
\end{teaserfigure}

\settopmatter{printacmref=false} 
\maketitle

\section{Introduction}
\input{01-introduction}

\section{Related Work}
\input{02-relatedwork}

\section{Task dimensions}
\input{03-taskdimensions}

\section{Tool capabilities}
\input{04-toolevals}

\section{Interaction Design and Implementation}
\input{05-implementation}

\section{User study}
\input{06-userstudy}

\section{Study Results}
\input{07-themes}

\section{Discussion}
\input{08-discussion}

\section{Conclusion}
\input{09-conclusion}

\section*{Disclosure}
The authors used GitHub Copilot v1.111.404 for code prediction in the preparation of figure source code.

\begin{acks}
  We would like to thank Shm Garanganao Almeda, James Smith, and Matthew Beaudouin-Lafon for their valuable insights that contributed to the framing of this work.
\end{acks}

\bibliographystyle{ACM-Reference-Format}

\bibliography{omnifill-citations.bib}

\end{document}

%% file: 01-introduction.tex
A significant but tedious part of information work consists of taking textual information from one source, and manually moving it into structured online forms - either through direct copy and paste, or through reformatting or rewriting.
Examples include extracting information from a purchase receipt into a reimbursement form (see Figure 1), adding events from the Web to a personal calendar, or transforming the format of existing form field values in-place. 
In these tasks, users are asked to laboriously serve as human ``glue'', often between different siloed systems.
Prior research has identified this problem and proposed several approaches to offer contextually-relevant form completions~\cite{stylos2004citrine,belgacem2023,hartmann2009,wang2013}.
One limitation of these prior approaches is that they often only apply in narrowly defined domains.
In particular, the choice of implementation approach, e.g., programming-by-demonstration~\cite{little2007koala} or document mining~\cite{bijalwan2014knn} enables some tasks, but precludes others.

In this paper, we first contribute a framework for characterizing the form filling task.
We describe four major dimensions -- information demands, operation complexity, structure variability, and task specification.
This analysis yields unique novel opportunities for LLMs to produce form filling suggestions for a wide range of tasks that were out of reach with prior approaches. 
In particular, we posit that LLMs are suitable for operating on a multi-faceted ``bag of context'' that can contain both prior demonstrations (as in programming by demonstration), as well as explicitly marked content (text flagged by the user as important) and implicitly collected content (history such as search queries and user behavior in the form).

We explore this research hypothesis with OmniFill, a new prototype system that collects multi-faceted context including browsing and text entry activity; constructs an LLM prompt with that content; and then offers suggestions \textit{in situ} for arbitrary structured form interfaces.
We demonstrate the promise of this approach through several example applications, and we examine utility through a study with 18 participants.
Through an analysis of users' behaviors and attitudes during the user study tasks, we propose considerations that system designers should keep in mind when building this type of domain-agnostic suggestion system for use in the real world.

%% file: 02-relatedwork.tex
The most closely related prior work falls into three areas: 1) automating interaction with Web pages in general; 2) form-filling interfaces in particular; and 3) other predictive text tools.
In addition, OmniFill is related to a rapidly growing set of user interfaces that use LLMs as backends.
We discuss each in turn.

\subsection{Automating Interactions with Web Pages}

Research has investigated several ways for automating tedious or repetitive interactions in Web pages beyond suggesting or filling forms.

One line of work uses end-user programming to record demonstrations of Web page interactions, generalize these demonstrations into a program in a suitable DSL, and then let others replay this program later in a related but different context.
Key tasks enabled by programming-by-demonstration have been sharing multi-step processes (e.g., Koala~\cite{little2007koala}, later CoScripter~\cite{leshed2008coscripter})
and Web scraping (e.g., Ringer~\cite{barman2016ringer}, Rousillon~\cite{chasins2018rousillon}).
A source of complexity in these systems is that Web pages may be interactive, their structure may be ill-formed, and they may change over time; more recent work has succeeded in using natural language processing techniques to make task specifications more robust and flexible (e.g. DiLogics~\cite{pu2023dilogics}).

Some other approaches to automation rely on pixel-based reverse engineering of rendered interfaces (Prefab~\cite{dixon2010prefab}, Sikuli~\cite{yeh2009sikuli}), applying computer vision techniques to derive interface structure.
The combination of DOM and visual features can surpass the limitations of these individual approaches~\cite{2013-webzeitgeist}.

OmniFill is not addressing general Web automation, focusing instead only the narrower problem of suggesting text in Web form fields.
However, it can draw on broader context than just prior demonstrations.
While demonstrations are part of the context considered by OmniFill, the system also takes into account other sources of context explicitly and implicitly collected (e.g., text identified as important by the user), and world knowledge (as captured in the pre-trained LLM).

\subsection{Automatic Form Filling}

OmniFill is inspired by prior mixed initiave systems such as LookOut~\cite{horvitz1999} and Citrine~\cite{stylos2004citrine}, which also seek to reduce the tedium of manually completing forms with information that already exists in some other format.
LookOut extracted information from email to pre-populate calendar events.
Citrine parsed users' copied text into typed fields using hand-constructed parsers for frequent content such as addresses, and could then auto-complete different address fields based on recognizing the structure of the form from prior demonstrations.
Related interaction techniques such as Entity Quick Click also rely on entity recognition in copied text to accelerate copy and paste tasks~\cite{bier2006entity}.
OmniFill extends this work by using an LLM and multiple sources of context to broaden the applicability of Citrine's approach without requiring manually authored parsers for each type of data.

A number of technical approaches have been proposed to extract information from a larger document or corpus for the purpose of form filling, e.g. hidden Markov models~\cite{borkar2000automatically}, or discriminative context free grammars~\cite{viola2005cfg}.
Prior work using ML approaches has constructed direct mappings between observed user context and form fields using NLP techniques (e.g.~\cite{hartmann2009}) or constructed domain-specific models of form dependencies (e.g.~\cite{belgacem2023}).
Because foundation language models, as used in OmniFill, can operate across many tasks and domains~\cite{bommasani2022opportunities}, they hold the promise of potentially obviating domain-specific recognizers and expanding task specifications beyond field-by-field extractions of suggestions from context.

\subsection{Predictive Text Completion}
Form filling is a special case of the larger problem of predictive text completion which seeks to predict, given some context of existing text, what text a user is likely to enter next.

Predictive Text Completion has existed for a long time in code editors~\cite{teitelbaum1981cps}, search query interfaces~\cite{bast2006typeless,nandi2007assisted}, and predictive keyboards~\cite{koester1994modeling}.
More recently, language models have found widespread use completion suggestions in email composition~\cite{goodman2022}, and for larger code chunks~\cite{nguyen2022empirical}.
In addition to research on the underlying technologies, researchers are also studying the impact of the use of predictive text interfaces on productivity~\cite{ziegler2022, palin2019people}, on the content of text being produced~\cite{arnold2020predictive}, and on users' perceptions of their tasks~\cite{jakesch2023co}.
While we restrict our focus in this paper to form filling, our conceptual framework and our multi-faceted context structure can potentially be applicable for analyzing a broader set of predictive tasks.

\subsection{Novel Interfaces Enabled by LLMs} 
Researchers are increasingly investigating the utility of pre-trained language models for enabling new interactions. 
One key distinction is between systems that use LLMs to enable natural language based input on one hand; and systems that use LLMs as an enabling implementation technology for novel direct manipulation or other types of interactions.

As examples of the first category,
Wang et al. show that LLMs are promising for enabling conversational interactions with mobile user interfaces~\cite{wang2023enabling}, such as screen summarization and screen question asking.
Stylette~\cite{kim2022stylette} enables re-design of Web pages through natural language commands.
In the second category, user interactions are translated into appropriate prompts to a language model ``behind the scenes'' to offer better performance or novel capabilities beyond previous algorithmic approaches.
Examples are TaleBrush~\cite{chung2022talebrush}, which allows users to sketch story arcs to condition the co-creation of stories with LLMs, and SayCan~\cite{saycan2022arxiv}, which uses LLMs to map real-world robot affordances to relevant situations.

OmniFill does not directly expose text interactions with the underlying LLM, instead offering a user-facing interaction closer to traditional brower autofill.
Still, the use of an LLM as implementation technology enables the system to perform many natural language tasks that are implicitly used in form filling, such as extracting entities from context and making use of form field labels.

%% file: 03-taskdimensions.tex
\label{sec:task-dimensions}

A form-filling task is a related series of operations intended to gather, process, and enter information into a form.
Form-filling tasks can manifest in many shapes, and each task has its own multidimensional requirements necessary of a system that is able to automatically suggest or complete field values for the task.

To assist in articulating the space of tasks that may be handled effectively by an LLM-backed system, we consider four relevant dimensions to describe form-filling tasks: information demands, operation complexity, structure variability, and task specification.

\subsection{Information demands} \label{sec:information-demands}
What information is being placed into the form?
Where does it come from?
In a predictive system, what information may be transcribed, transformed, or used as a retrieval key for the system's current suggestion or completion?

We identify six broad classes of such information.
\begin{description}
    \item[Historical user behavior] What information has the user entered into the form or similar forms in the past? Example: browsers offer dropdown suggestions for recognized fields, even in previously-unseen forms.

    \item[Explicitly-foregrounded information] Users may explicitly call attention to information outside the target form interface. Example: importing a CSV into a system, or the ``select, then copy'' operation of a clipboard.

    \item[Implicit browsing context] User activity may provide additional context relevant to the task even if not explicitly foregrounded. Example: scrolling through the subject lines of an email inbox.

    \item[Current form state] Some form filling operations make use of information already present in the form. Example: creating a username based on the values of a ``First name'' and ``Last name'' field.

    \item[External general knowledge or language knowledge] Some information inserted into forms comes from the surrounding world, especially in combination with other information sources. Example: after typing ``Paris'' in a ``City'' field, a system may offer ``France'' for the ``Country'' field.

    \item[User-specific external knowledge] User-specific information may come from sources other than form-filling activity or recent browsing context. Example: a system may offer completions based on the data from the user's phone contacts.

\end{description}

\subsection{Operation complexity}
What transformations must be applied to inputted information by either the user or a predictive system making field suggestions?

\begin{figure*}
    \centering
    \includegraphics[width=\textwidth]{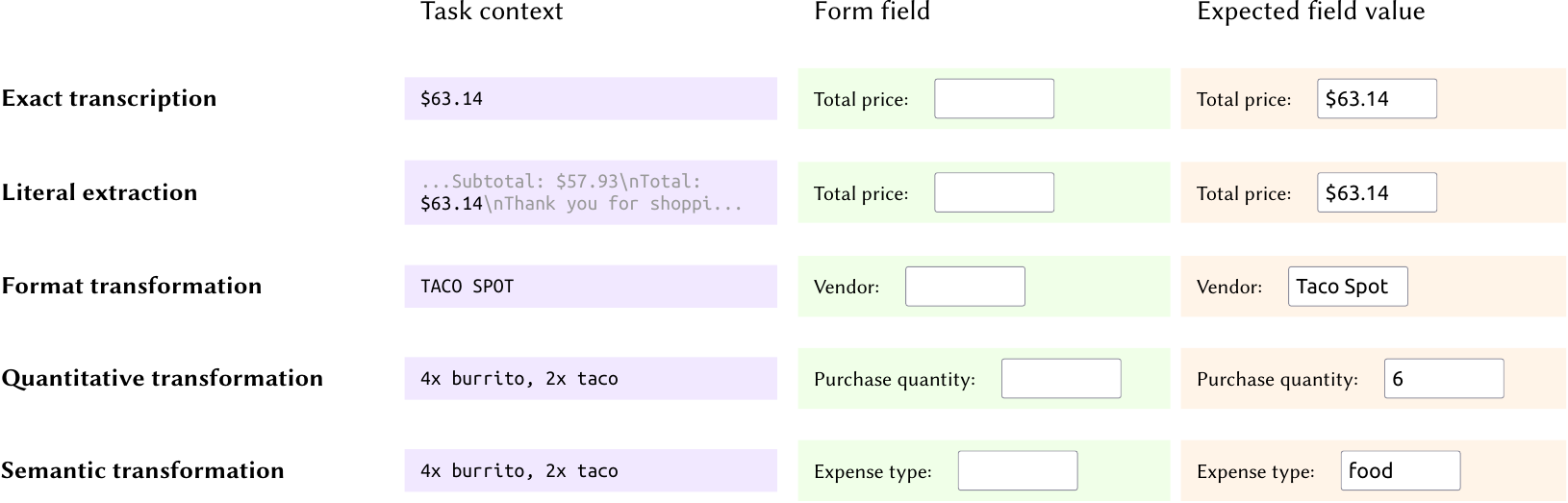}
    \caption{Illustrations of various categories along the operation complexity dimension.}
    \Description{Examples of operation complexity, as described in the associated section, are shown in a table.}
    \label{fig:operation-complexity}
\end{figure*}

\begin{description}
    \item[Exact transcription] An entire input is transcribed verbatim into the output field. Example: a user copies a URL into their clipboard, and their mobile browser address bar offers to ``paste'' that full URL.

    \item[Literal extraction] Information (in any of the forms described in Section \ref{sec:information-demands}) is transcribed verbatim into the target form but must first be extracted from some larger source.

    \item[Format transformation] Information must be transformed into a different format. Example: a task requires a name to be converted from all-caps (as it appears in the source) to title case (as required by the target form).

    \item[Quantitative transformation] The task demands arithmetic operations to be performed on available information before inserting into the target form.

    \item[Semantic transformation] The task requires a transformation of the available information that produces an output that is syntactically and structurally distinct from the input information but still semantically related. Example: upon viewing an email reading, ``I don't eat meat, but I can eat animal products like cheese'', the user might write ``Vegetarian'' into the target form to match the format of other form responses.
\end{description}

\subsection{Structure variability}
When the task consists of multiple instances of a form being filled, how does the task vary between form fills?
Once the task is well-specified, how flexible does that specification need to be to handle the source and target structures?

\begin{description}
    \item[Fixed source structure] Does information used to derive field values come from the same place, in the same format, from case to case? An example of a task with \textit{fixed} information source structure is one in which information is sourced from a spreadsheet and each row of the sheet is used to construct a submission in the target form.

    \item[Varied source structure] An example task with \textit{varied} information source structure is the information-gathering task from our user study, in which users repeatedly fill the same form but with information from a differently-structured website for each submission of the target form.

    \item[Fixed target structure] Is information always inserted into the same field of the same form, or can each submission to a form take a different structure? Any task involving repeated submissions to the same form has a \textit{fixed} target field structure.

    \item[Varied target structure] An example of a task with \textit{varied} target field structure is filling out many distinct job applications; although the information being filled is mostly the same from submission to submission, the form structure (and in some cases, the demanded information format) changes each time a form is filled.
\end{description}

\subsection{Task specification}
What information is available that specifies the process of completing the task?
If a user is completing the task manually, how much understanding of the task might be gleaned by someone looking over their shoulder? These specifications are distinct from the information demands described in Section \ref{sec:information-demands}, which refers to information that may be actually entered into the form (potentially after a transformation) or used to reference other information to be entered into the form.

\begin{description}
    \item[Implicit specification] Task specifications may be implied by the structure of collected context or target form. For example, a user who copies an address and visits a form with an address field likely intends to insert that address into the target form. This structural information may take a machine-readable form (as with the \texttt{autocomplete} attribute in HTML) or natural language forms (as with text description labels on fields).

    \item[Instructions] Tasks can be specified with form-filling instructions, present in, for example, the form or browsing context. Instructions may be precise and machine-readable, as with tasks specified using code or macro software. They may also come in the form of natural language, which may leave gaps in the specification.

    \item[Examples] Many tasks can be approximately specified using a number of examples, as demonstrated in the programming-by-example literature. When prior examples of user behavior are available, approximate task specifications can be inferred and realized in synthesized code. Examples may also help to clarify ambiguities in vague instruction-based specifications.
\end{description}

\subsubsection{Specification visibility and mutability}
In a predictive system offering form-filling suggestions, allowing the user to see and manipulate the specification can enable refinement to improve future suggestions.
This can happen by manually constructing or deleting examples, or by modifying instructions, when they are present and in a form that the user is comfortable manipulating.
Many programming-by-example techniques produce a readable or manipulable code specification as an artifact, rather than keeping the specification as a hidden implementation detail~\cite{chasins2018rousillon, cambronero2023flashfill}.

Specification refinement need not be driven by the user completing the task.
If predictive systems can detect incomplete specifications or present the user with anomalous saved examples, they can take initiative in asking the user to clarify the task specification (as discussed in~\cite{horvitz1999, mayer2015user}), just as another person looking over the user's shoulder might.

%% file: 04-toolevals.tex
\begin{figure*}
    \centering
    \begin{tabularx}{\textwidth}{|
    >{\raggedright\arraybackslash}p{.1\textwidth}|
    >{\raggedright\arraybackslash}p{.15\textwidth}|
    >{\raggedright\arraybackslash}X|
    >{\raggedright\arraybackslash}X|
    >{\raggedright\arraybackslash}p{.15\textwidth}|
    >{\raggedright\arraybackslash}p{.2\textwidth}|}                                    \hline
                                    & \textbf{Browser autofill} & \textbf{LookOut} & \textbf{Citrine} & \textbf{DiLogics} & \textbf{OmniFill}   \\ \hline
    \textbf{Information demands}        & Historical user behavior, Current form state
                                        & Implicit browsing context
                                        & Explicitly-foregrounded information
                                        & Historical user behavior, Explicitly-foregrounded information
                                        & Historical user behavior, Explicitly-foregrounded information, Implicit browsing context, Current form state, External general knowledge or language knowledge \\ \hline

    \textbf{Operation complexity}       & Exact transcription
                                        & Exact transcription, Literal extraction, Format transformation
                                        & Exact transcription, Literal extraction, Format transformation
                                        & Exact transcription, Literal extraction, Format transformation 
                                        & Exact transcription, Literal extraction, Format transformation, Semantic transformation \\ \hline

    \textbf{Structure variability}      & Fixed source structure, varied target structure
                                        & Fixed source structure, fixed target structure
                                        & Varied source structure (only in known schemas), varied target structure
                                        & Fixed source structure (data import), varied target structure
                                        & Varied source structure, varied target structure  \\ \hline

    \textbf{Task specification}         & \textbf{N/A} (fixed to tasks handled by the system)
                                        & \textbf{N/A} (fixed to tasks handled by the system)
                                        & \textbf{N/A} (fixed to tasks handled by the system)
                                        & Instructions (natural language), Synthesized instructions (code), Examples, Visible/mutable specification
                                        & Implicit specification (natural language), Instructions (natural language), Examples \\ \hline
    \end{tabularx}
    \caption{Task requirements supported by browser autofill, LookOut, Citrine, DiLogics, and OmniFill.}
    \label{fig:tool-eval-table}
\end{figure*}

To situate OmniFill among prior form-filling tools, we analyze the capabilities of existing tools with respect to the space of tasks they are capable of handling, as shown in Figure \ref{fig:tool-eval-table}.

First, we consider typical Web browser autofill behavior, which matches form fields to prior values entered into similar fields by the user.
Then, we compare to LookOut~\cite{horvitz1999}, a tool for Microsoft Outlook that offers automated scheduling services based on the contents of users' emails.
By comparing Citrine, an ``intelligent copy-and-paste'' tool~\cite{stylos2004citrine}, we introduce a tool that can detect information from certain known schemas in flexible information sources.
Then we consider DiLogics~\cite{pu2023dilogics}, a programming-by-demonstration tool that enables automated form filling from an imported CSV and synthesizes an explicit specification from existing examples.

Implementation approaches in this prior work constrain the types of task that can be performed.
First, each approach offers a structure that privileges some forms of information over others, rather than allowing information sources to vary based on the task demand.
Second, prior approaches cannot perform semantic transformations without domain-specific processors in place.
Many prior systems are also locked to repeating source or target structures, rather than supporting tasks with flexible needs.
Systems that require heavy specification, though valuable for full automation, cannot quickly make lightweight inferences of approximate details or implicit specifications in flexible contexts.

Pre-trained foundation language models may be uniquely suitable to address these shortcomings: as long as relevant context information can be brought into the system as text, different types of information can be mixed in a single prompt.
LLMs also offer semantic transformation capabilities, and the world knowledge captured by models during their pre-training may give them the power to adapt to arbitrary tasks seen by users.
We therefore built OmniFill with an LLM backend, seeking to cover many possible information demands, a range of operation complexity types, flexible source and target structure, and flexibly-defined task specifications.
Not every task is well-suited to LLMs (e.g. because of limited context window size or poor arithmetic skills), but even these shortcomings may be ameliorated using targeted techniques developed by AI researchers~\cite{lewis2021retrievalaugmented,wei2023chainofthought,schick2023toolformer}.

%% file: 05-implementation.tex
\begin{figure*}
    \centering
    \includegraphics[width=\textwidth]{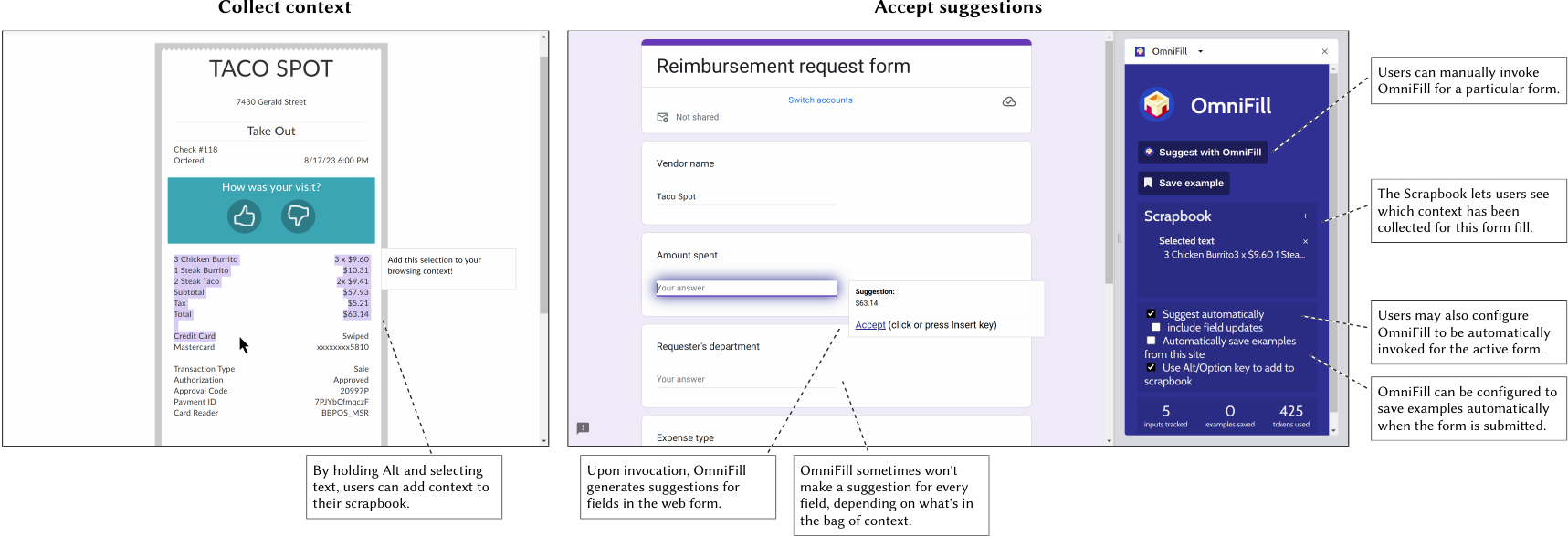}
    \caption{OmniFill's browser extension. Users can select text on the Web to add to their scrapbook, then invoke the system to obtain suggestions for any Web form. A sidebar on the right of the browser controls the scrapbook, invocation, and configuration.}
    \Description{On the left, labeled ``Collect context'', is a screenshot of a browser window containing a receipt for a restaurant called ``TACO SPOT'', with much of the receipt highlighted in purple. Next to the highlight is a box in the screen containing the text ``Add this selection to your browsing context!''. On the right, labeled ``Accept suggestions'', is a screenshot of a Google form labeled ``Reimbursement request form''. One field, ``Amount spent'', has an abutting box offering a fill suggestion, ``\$63.14''. On the right of this screenshot is a sidebar, containing buttons and checkboxes as labeled in the accessible text of the figure.}
    \label{fig:system-ss}
\end{figure*}

\subsection{System architecture}

\begin{figure}
    \centering
    \iftoggle{manuscript_version}{
        \includegraphics[width=0.5\textwidth]{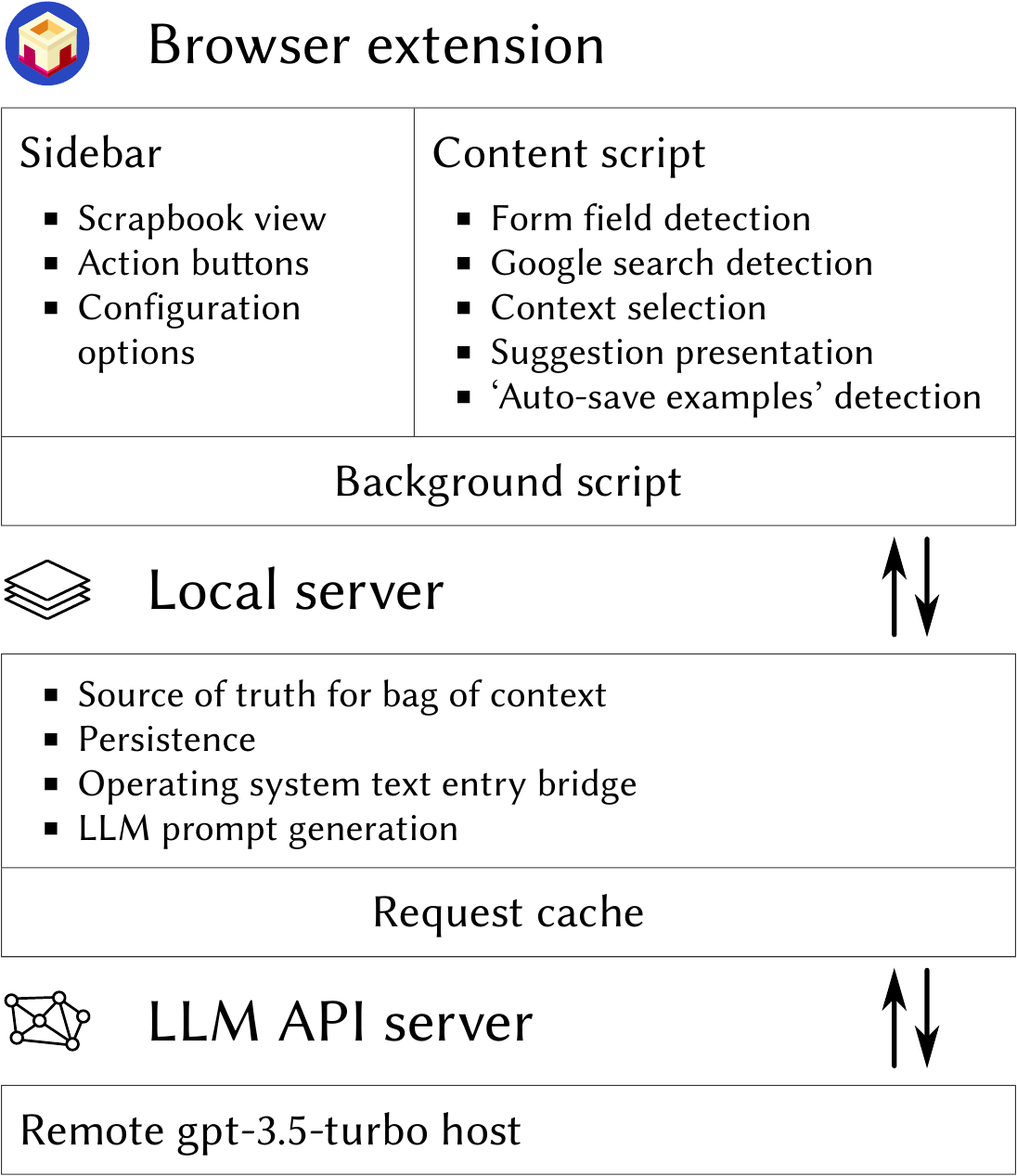} 
    }{
        \includegraphics[width=0.75\columnwidth]{system-architecture-crop.pdf} 
    }
    \caption{OmniFill's underlying architecture, including the browser extension, local server, and remote API server.}
    \Description{Three boxes, labeled ``Browser extension'', ``Local server'', and ``LLM API server'', stacked vertically, with arrows pointing in both directions between each adjacent pair of boxes in the stack. The text in the boxes is accessible.}
    \label{fig:sys-arch}
\end{figure}

OmniFill consists of two local components, a browser extension and local server both implemented in TypeScript, and a connection to a remote LLM server with a ``chat completion'' generation API.

Together, these components enable a streamlined interaction flow, depicted in the system screenshots in Figure \ref{fig:system-ss}, offering users low-friction interactions to foreground relevant context, invoke OmniFill's backend to offer completion suggestions, and show the system examples of the form-filling task.
This all occurs \textit{in situ}, requiring no task-specific configuration and minimal interaction outside the context sources and target form already in use.

\subsubsection{Browser extension}
OmniFill includes a sidebar as part of its Web browser extension, which contains a preview of the contents of the ``scrapbook.''
The scrapbook contains browsing context observed by OmniFill, including Google searches, text from Web pages added to the scrapbook, and any text manually added through the sidebar's add-to-scrapbook button (which presents a free-form text field that users can paste or type text into).
Users can delete individual scrapbook entries from the scrapbook but not view the full scrapbook contents or make in-place updates to collected context.

The sidebar also includes two action buttons: ``Suggest with OmniFill'', which invokes the LLM to make predictive suggestions for the currently-focused Web page, and ``Save example'', which captures the current scrapbook contents as well as the current page's form structure to save into the ``Prior examples'' section of OmniFill's bag of context, then clears the scrapbook in preparation for the next form-filling example.

The sidebar also contains configuration options such as OmniFill's automatic invocation mode (which can update suggestions when the scrapbook changes or form fields are modified) and its automatic example-saving feature, which attempts to detect form submissions and saves examples automatically.

OmniFill's browser extension also injects JavaScript into each visited Web page, responsible for collecting context from the Web and from user activity.
Users may select text while holding the Alt (or Option) key on their keyboard, which adds the selected text to OmniFill's bag of context.
In addition, Google search queries are automatically added to the context.
The injected script also synchronizes the current page's form field structure, including initial field values and updates made by the user to the fields, to the rest of the system.
Form fields have their names inferred through a variety of methods; first, OmniFill attempts to read the fields' accessible names (e.g. \texttt{aria-label}s), falling back to nearby visible text if necessary.
If the automatic example-saving feature is enabled for a particular site, this injected script listens for form submissions or clicks on buttons labeled ``Save'' or ``Submit'' and invokes the example-saving routine.
This technique may not generalize to all websites, so users may need to click the ``Save example'' button manually before making form submissions;
in our user study, this technique was sufficient to save examples automatically.

When OmniFill has suggestions for the user, the injected script seeks out the associated form fields in the page by their inferred name, highlights them with a purple outline, and offers a small suggestion box when focusing fields whose suggested value differs from the current value.

\subsubsection{Local server}
OmniFill's browser extension coordinates with a local server program running on the computer.
This separate server simulates key presses when OmniFill needs to type into a text field. 
In addition, this server acts as the single source of truth for OmniFill's current bag of context, and this is persisted to the user's computer in case the browser or server is closed.
When OmniFill is ready to make a request to the LLM, this local server generates a prompt, computing the number of tokens used by the prompt and pruning examples and scrapbook contents if necessary until the prompt is sufficiently short to fit in the model's context window.
Responses are cached by the local server so that only unique requests are made to the LLM API.

\subsubsection{LLM API}
\begin{sloppypar}
OmniFill makes requests to OpenAI's \texttt{gpt-3.5-turbo-0613} model, choosing between the 4,096-token context length model and the 16,384-token model based on the size of the prompt.
The model is queried at temperature 0, offering nearly deterministic results for a particular prompt.
Responses are piped from the local server back to the browser extension, which presents any suggestions back to the user.
\end{sloppypar}

\subsection{Serializing the ``bag of context'' into an LLM prompt} 
\begin{figure}
    \centering
    \iftoggle{manuscript_version}{
        \includegraphics[width=0.4\textwidth]{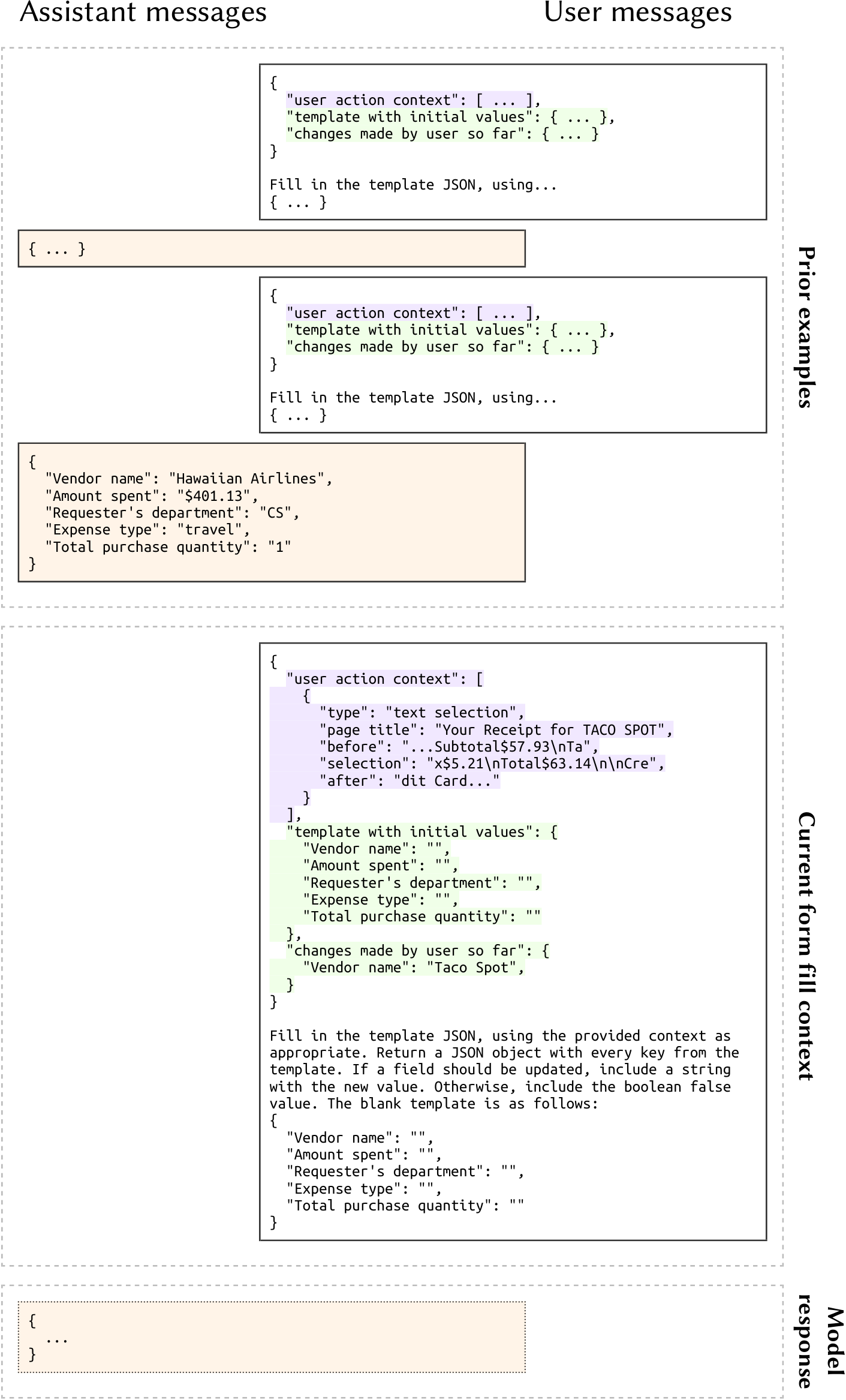} 
    }{
        \includegraphics[width=\columnwidth]{prompt-structure-figure-fixed-nobacktick-crop.pdf} 
    }
    \caption{The structure of the prompt generated by a single invocation of OmniFill for \texttt{gpt-3.5-turbo}. Some details are omitted from prior examples. In addition, OmniFill expands selection context by 500 characters in each direction. Highlighted in purple are details about the browsing context (Section \ref{sec:prompt-browsing-context}); in green is form structure state (Section \ref{sec:prompt-form-structure-state}), and in orange is the output format (Section \ref{sec:prompt-output-format}).}
    \Description{A number of JSON objects are shown. The boxes to the right, each containing ``user action context'' and other fields, are labeled ``User messages''. The remaining JSON boxes are to the left of the figure. The top four boxes are labeled ``Prior examples'', the fifth box is labeled ``Current prompt'', and the final box is labeled ``Model response''.}
    \label{fig:prompt-structure}
\end{figure}

To allow the system to make useful inferences in varied situations, OmniFill constructs a prompt containing three main facets of task context (visualized in Figure \ref{fig:prompt-structure}):
\begin{itemize}
    \item \textbf{Current browsing context}, representing context explicitly inserted into OmniFill's scrapbook by the user as well as implicitly-observed Google searches.
    \item \textbf{Form structure state}, including the inferred name for each form field, the initial value of each form field, and any edits made by the user to each form field's value.
    \item \textbf{Prior examples} saved by the user, each including browsing context, form structure, and the intended final form state before saving the form.
\end{itemize}

This prompt structure is designed to support many broad tasks without requiring task-specific configuration.
Some tasks may primarily use only one or two of these context facets, and others may combine information from all three.
Depending on the task, these context facets may all contain \textit{information demands} of the task, and they may all contain information that helps OmniFill infer and carry out a \textit{task specification}, so all three facets are included in each prompt to the LLM. 

\subsubsection{Current browsing context} \label{sec:prompt-browsing-context}

All information present in the ``scrapbook'' in OmniFill's sidebar will be included in this section of the prompt, referred to internally as \texttt{"user action context"}.
This includes a description of Google searches made by the user, context added manually through the sidebar's ``add context'' button, and context added by selecting text on Web pages.
When text is selected on Web pages, 500 characters before and after the user's selection are also included in the prompt (separated, so that the prompt includes information about which text was actually selected).
In addition, the page's title (as shown in the browser tab header) is included.

\subsubsection{Form structure state} \label{sec:prompt-form-structure-state}

For the LLM to return results in a usable format, the prompt needs to contain a full list of the text fields on the target Web page.
The prompt also includes the initial value of each field and a list of the field updates that the user has made since those initial values were retrieved; this provides context on how the user has engaged with the form so far, which can, for example, allow the system to make suggestions for the second half of a form after the user has given the system indications of their intent through edits made to the first half of the form.
The ``initial'' value of each field is computed either at page load time or the time the form structure last changed, whichever came later -- this enables the prompt to ``start fresh'' in situations where form submission does not reload the page but instead adds new fields to the page (e.g. when clicking a button to add a new row to a spreadsheet-style form).

In practice, we discovered that, if some fields had already been updated by the user, the system sometimes would not offer suggestions for the remainder of the form, instead making only suggestions that affirmed what the user had already input.
In the OmniFill prototype, we make two parallel requests to the model: one request as described in this section, and a secondary request with any current user edits suppressed (but the initial field values still present), using responses from the second prompt only as a fallback when a field did not have a suggestion from the first prompt.
In the information-gathering task of our user study (described in section \ref{sec:task-1}), 38 of 418 (= 9\%) of model completions had at least one field suggestion made by this metric that would have been left blank or unchanged if not for this secondary parallel request. 

\subsubsection{Prior examples}
Because \texttt{gpt-3.5-turbo} is a ``chat completion'' model, we can model prior examples as previous ``exchanges'' with the model by including past prompts (as described in the prior two subsections) as a ``user message'' and the desired output of the LLM (the example's final form state, as saved by the user) as an ``assistant response'', including final values for each field in the response.
The prompt does not include actual prior model responses for previous requests, only the final field values saved by the user in the ``assistant response''.
By saving examples, especially when filling the same form repeatedly, users can demonstrate their process to OmniFill.
Even when OmniFill does not successfully extract meaningful names from the fields in the target form, we have observed that a small number of examples can suffice to begin specifying the form-filling task.

\subsubsection{Output format} \label{sec:prompt-output-format}

A simple structure for the requested output format (which is duplicated in the prompt's ``assistant message'' for each example) is a JSON object with a key for every field in the target Web form and a value equaling the system's suggestion for that field.
By requesting suggestions for the entire form in one query (rather than requesting each field's suggestion separately, in parallel), the system can produce internally-consistent suggestions, which may be valuable in some tasks where there are multiple reasonable completions for the full form.

After some experimentation, we settled on an output format that asks the model to respond with a JSON object that includes every key, but to include string values only for the fields that should have their values updated from their current values.
Otherwise, the JSON value for this field should be a \texttt{false} Boolean value.
By keeping each key in the JSON object (as opposed to asking the model to return only the keys needed), we observed a lower risk of the model prematurely closing the JSON object in tasks where the model needed to make many suggestions; this serves as a rudimentary ``chain-of-thought'' prompt~\cite{wei2023chainofthought}, considering each key individually before making a decision to finish the response.
By removing the requirement to transcribe many fields exactly, we prevent a distracting, noisy set of output examples from obfuscating the true task being demonstrated in those examples.

We have found that, given the prompt as described in this section, OmniFill rarely suggests overeager fills for form fields that (given the bag of context) don't need to be updated in the tasks for which we have tried the system.
In our user study's information-gathering task, for example, OmniFill rarely provided a suggestion for a field whose value was not discernible from browsing context, and in the data-formatting task, OmniFill typically only offered suggestions for the fields that the task asked users to update.
The extent to which OmniFill is conservative this way is highly dependent on the prompt; for example, altering the prompt in Figure \ref{fig:prompt-structure} to include the sentence, ``You must provide a useful suggestion for every field, even if you aren't sure.'' causes the model to offer a suggestion for every field.

We believe that, although LLMs can be powerful engines for solving tasks in a domain-agnostic way, tuning prompts to improve performance on some tasks without hindering performance on other tasks requires careful planning.
To better manage the uncertainty of ``herding AI cats''~\cite{zamfi2023herding}, practitioners should regression-test changes to systems' prompt structures on a wide range of sample tasks that require capabilities across all task dimensions (as described in Section \ref{sec:task-dimensions}) and context facets.

%% file: 06-userstudy.tex
We conducted one-hour sessions with 18 participants to judge impressions of OmniFill. Participants, aged 21-30 (mean 24), spoke English and were located in the United States.
Four participants identified as female, twelve identified as male, and two did not disclose their gender.
Participants completed a consent form and were compensated with a \$20 gift card for their time.

\subsection{Information-gathering task} \label{sec:task-1}
\begin{figure}
    \centering
    \iftoggle{manuscript_version}{
        \includegraphics[width=\textwidth]{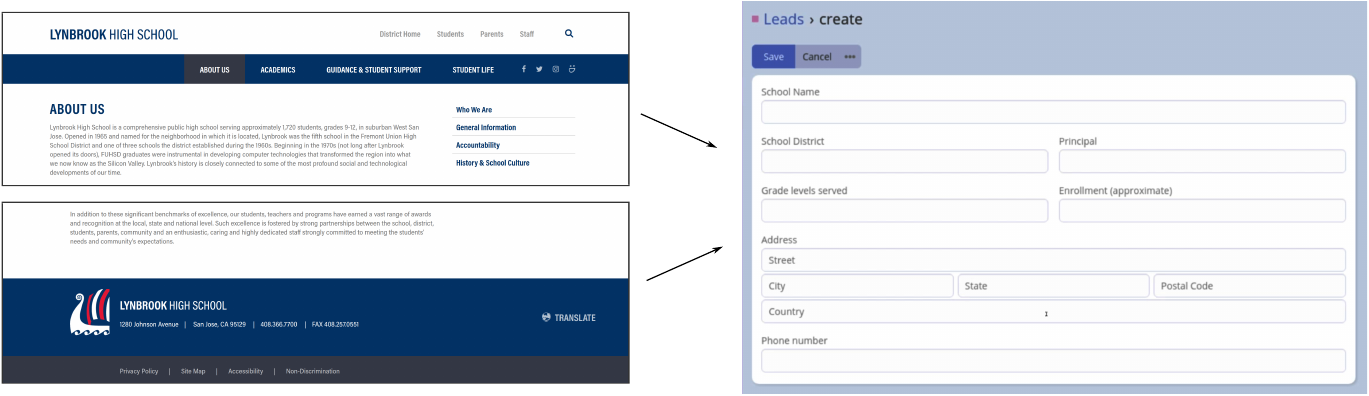} 
    }{
        \includegraphics[width=\columnwidth]{task-1-ss-crop.pdf}  
    }
    \caption{In the information-gathering task, information from school websites is inserted into the target form in the information-gathering task, including fields for: school and district name, principal name, grade levels served (e.g. 9-12), total enrollment count, address, and phone number.}
    \Description{A screenshot of the target form in the information-gathering task, including fields for: school and district name, principal name, grade levels served (e.g. 9-12), total enrollment count, address, and phone number.}
    \label{fig:lead-gen-form}
\end{figure}

Participants joined a Zoom call and completed two distinct tasks with OmniFill through a remote virtual desktop, with a demonstration of the system incorporated into the first task.
In the first task, participants were instructed to ``forage'' information~\cite{pirolli1999information} from websites of schools, inputting that information into a structured form (an instance of EspoCRM~\cite{espocrm}), shown in Figure \ref{fig:lead-gen-form}.
First, participants were directed to the website of a public high school and asked to spend a few minutes to fill the form manually, without using OmniFill.
For this test website, participants who took longer than a few minutes were offered assistance locating the form data.

Then, participants were shown a demonstration on a different school's website, using OmniFill to fill the same form.
In this demo, participants were shown how to use the Alt+select interaction to add context to their scrapbook, then shown how to generate and accept suggestions for the form.
Participants were shown that selections can be made approximately, and that it was possible (but not necessary) to make multiple distinct selections before returning to the target form.
Participants were shown how to add context to the scrapbook manually, for situations where the Alt+select interaction failed (e.g. when viewing PDFs).
In the demonstration, participants were shown and reminded that form inputs may be typed manually and need not come from OmniFill suggestions.
We also checked OmniFill's ``Automatically save examples for this site'' checkbox so that the form's ``Save'' button would trigger the current form inputs to be saved as an example.

After the demonstration, participants were asked to practice using the system by again filling the form with data from the website they had already manually collected information from.
Then, participants were given approximately 15 minutes to perform the task freeform, finding websites for other schools and filling out the form, one school at a time.
To encourage participants to find websites with varying structures, we asked them to research schools from different school districts.
We also asked participants not to spend more than a few minutes on any one website and indicated that it was okay to leave some fields blank if they didn't think they would find the information.

\subsection{Data formatting task}
\begin{figure}
    \centering
    \includegraphics[width=\columnwidth]{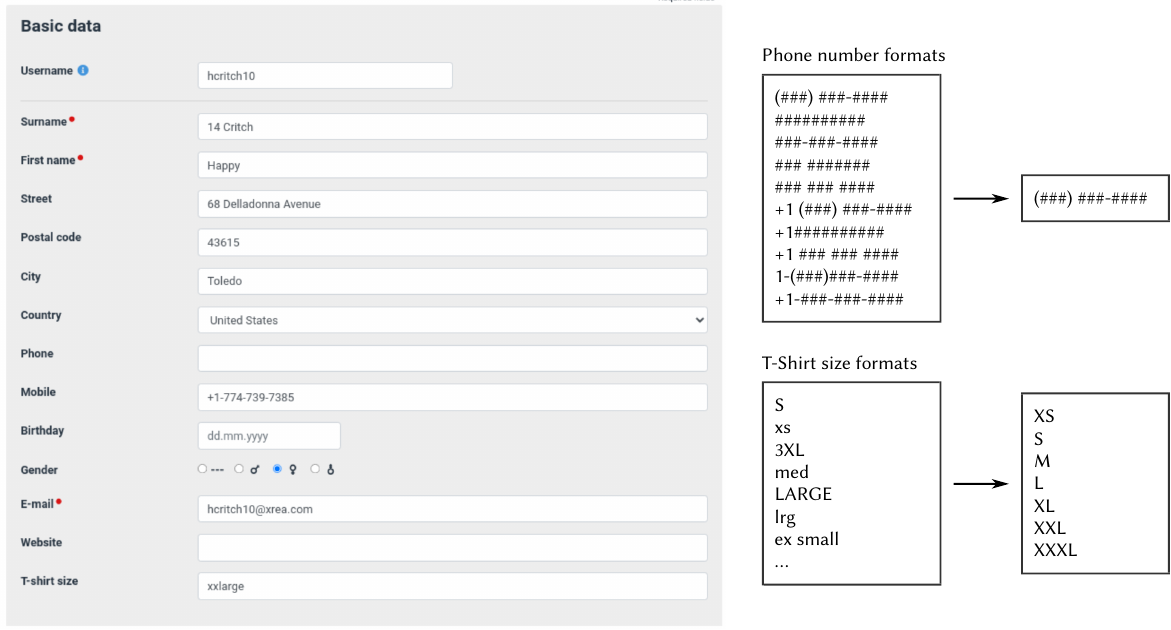}
    \caption{The form and required transformations for the data-formatting task.}
    \Description{A long form with pre-filled synthetic company member data, notably including a ``Phone'' field, a ``Mobile'' field, and a ``T-shirt size'' field. To the right, a depiction of the data-formatting task, including a reformatting from many possible phone number formats into one correct format and a reformatting of the T-shirt size.}
    \label{fig:admidio}
\end{figure}

The data-formatting task required OmniFill to recognize patterns when editing forms with prepopulated data.

Participants were shown a mock ``human resources'' website (an instance of Admidio~\cite{admidio}), pre-filled with fake company membership data. 
We prepopulated every user profile with a T-shirt size, most profiles (45 of 51) contained a mobile number, and some (14 of 51) contained a home phone number.
The string format of the shirt size and phone number were randomly chosen for each member profile, as depicted in Figure \ref{fig:admidio}.
Each participant received the same mock company data, but the order of the data was randomized so that OmniFill would behave differently for each participant.
No participant saw more than 28 of the 51 total member profiles during the ten minutes allotted to perform the task.

Participants were shown a series of instructions for their task, asking them to update each phone number in the member profile (if present) to a fixed format and to update each T-shirt size to one of a few fixed options (depicted in Figure \ref{fig:admidio}).
These instructions were shown in a screenshot so that participants could not Alt+select them, since we wanted to investigate OmniFill's pattern recognition; although participants could have manually typed the instructions into the scrapbook, none did.
Then, we instructed participants to visit each member profile in order, updating the phone number and T-shirt fields following these instructions, with OmniFill automatically saving examples each time the form was saved.
We also instructed participants to click the ``Suggest with OmniFill'' button in the extension sidebar and wait for its response on each profile before making any edits, testing the system's ability to recognize the task specification and offer the correct edits as suggestions.
We advised participants that they were not required to accept OmniFill's suggestions or use its suggestions unchanged.

\subsection{Interview}

After completing both tasks, participants were asked briefly about their experience in a conversational interview, including questions about OmniFill's perceived utility, concerns about real-world use, confidence in the system's accuracy, and confidence that they would notice OmniFill's mistakes.

%% file: 07-themes.tex
We first discuss high-level impressions for the two tasks, then introduce four important lenses through which we consider users' behavior; opportunistic scrapbooking, trust in OmniFill's suggestions, value in partial success, and a need for visibility into prompt context.

\subsection{Impressions}

Participants readily used OmniFill in the information-gathering task, accepting OmniFill's suggestions for almost all form fields in each school contact they saved; seventeen participants, when asked, said they would likely use OmniFill if they had to perform the information-gathering task in real life.
Although suggestions for this task largely used information from the \textit{current browsing context} portion of the prompt, even this task was able to benefit from OmniFill's multi-faceted prompt.
For example, participants who typed schools' country name as ``United States'' tended to receive this as a suggestion, where participants who preferred ``USA'' often received the suggestion in this format, indicating that OmniFill was making use of prior examples in its prompt.

\begin{figure*}
    \centering
    \includegraphics[width=\textwidth]{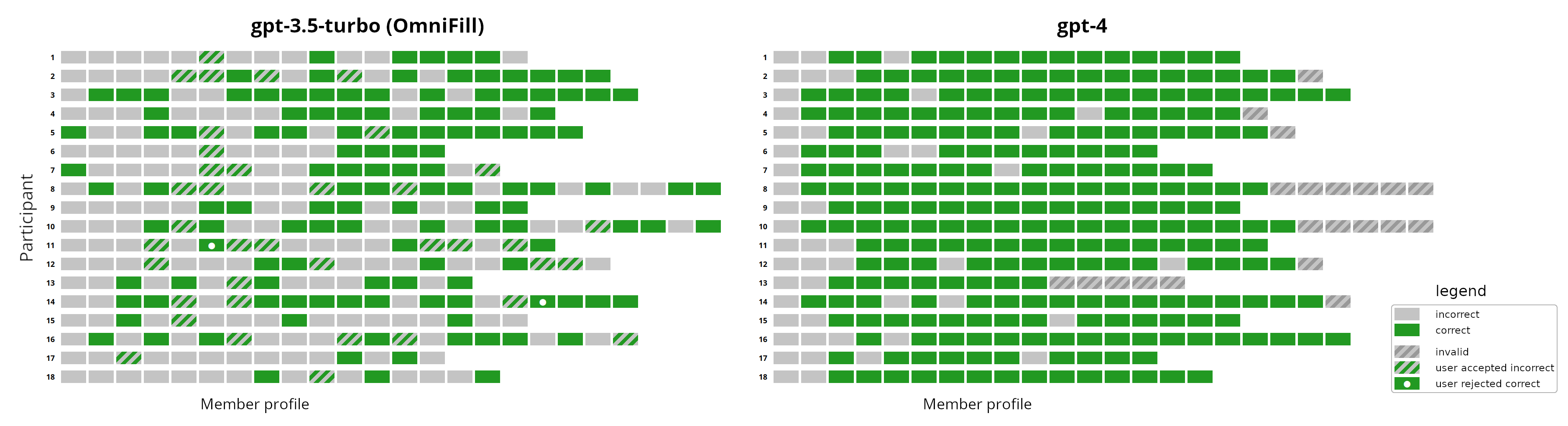}
    \caption{OmniFill's success in suggesting phone number formats in the data-formatting task, compared to those same prompts run against the GPT-4 API. Prompts that could not be run with GPT-4 due to context window restrictions are shown as ``invalid''. Each row represents one participant's experience in the task, showing suggestions left-to-right as the task progressed.}
    \Description{A figure depicting which requests in the data-formatting task contained the correct phone number information when OmniFill returned results. Successes are sparse, but for most participants, success rate increases (to around 60\%) towards the right of the figure. Most participants are shown to have accepted incorrect suggestions at least once. To the right, a figure depicting the same requests as made to GPT-4 shows a much higher success rate; though most requests fail within the first two or three columns, success rate quickly reaches a number closer to 90\%.}
    \label{fig:task-2-success}
\end{figure*}

In the data-formatting task, OmniFill proved more successful for some participants than for others, and when the system did eventually succeed more consistently, many examples were often required for this to settle.
Figure \ref{fig:task-2-success} visualizes OmniFill's success suggesting values for the phone number fields in particular.
Each row represents a participant, and each box represents a completion request sent through the system.
Only the first request is shown per member profile, even if participants invoked the system more than once, and we do not show requests for profiles where no phone number change needed to be made.
In this figure, the box is shown in green only when \textit{both} phone number fields are suggested correctly (although in many cases there was only one phone number, so only one change needed to be made).
Cases where users accepted an incorrect suggestion or rejected a correct suggestion (choosing instead to type the phone number manually) are also shown.

After conducting the study, we also called the GPT-4 API~\cite{openai2023gpt4} using the same prompts as those used in the study to observe system performance with the more powerful model, shown on the right side of Figure \ref{fig:task-2-success}.
Because GPT-4 has a smaller context window (8,192 tokens) than \texttt{gpt-3.5-turbo-16k}, some prompts could not be run for this analysis; those are depicted with gray stripes.
In addition to the context window size, real-world trade-offs of using GPT-4 for OmniFill include cost and latency; for these reasons and to better understand users' behavior with a less-effective model, we used GPT-3.5 for our studies.

\subsection{Opportunistic scrapbooking}

Compared to the meticulous copying-and-pasting (or memorizing-then-typing) strategy participants employed to fill out the information-gathering task's contact form before being introduced to OmniFill, participants' information-foraging behavior during the free-form component of this task was approximate and opportunistic.
We observed participants collecting information for OmniFill's bag of context quickly and even \textit{haphazardly}, as one participant put it, often with little regard for the later extraction step.

\begin{figure*}
    \centering
    \includegraphics[width=\textwidth]{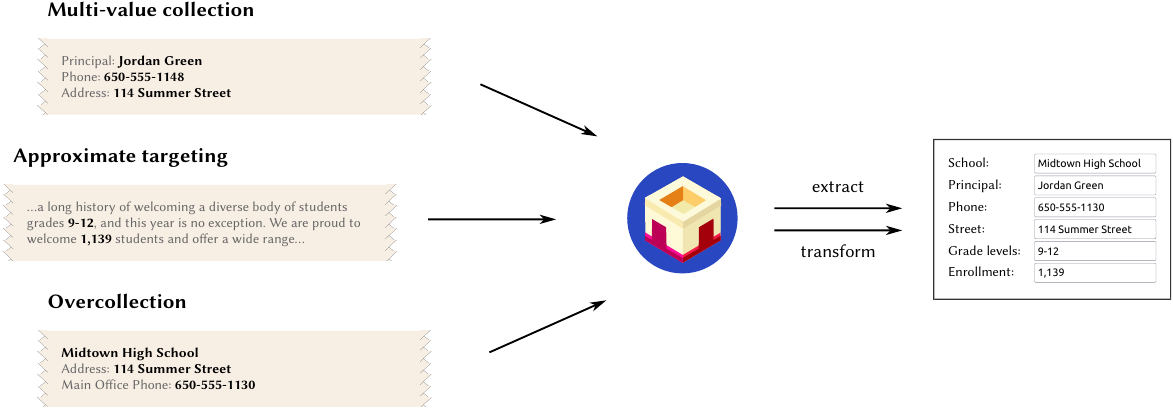}
    \caption{Opportunistic scrapbooking involves the collection of many values at once, approximate relevance judgments, and overcollection. OmniFill often extracts appropriate information even from opportunistically-built scrapbooks.}
    \Description{On the left, three pieces of collected context are shown. The first demonstrates multiple values being collected at once. The second demonstrates approximate targeting by showing information embedded in a larger paragraph. The third shows ``overcollection'', in which information is collected more than once, resulting in multiple phone numbers being present in context. All context flows through OmniFill, then through extraction and transformation to the final form suggestion.}
    \label{fig:haphazard-scrapbooking}
\end{figure*}

\subsubsection{``Collect-then-fill'' strategy}
\begin{sloppypar}
In the manual portion of the information-gathering task, participants all alternated frequently between the information source website and the target interface (the contact form), copying or transcribing information in small increments.
Because the contact form's ``address'' field was split into five text fields (``Address'', ``City'', ``State'', ``Postal Code'', and ``Country''), two participants constructed an ad hoc ``scrapbook'' inside the ``Address'' field, pasting a full line (address, city, state, and postal code) into the interface and then manually organizing it piece-by-piece into the appropriate fields.
Still, all participants were able to search only for small portions of the total information demand at a time before needing to insert the collected data into the contact form.
\end{sloppypar}

However, in the second phase of the information-gathering task, during which participants were equipped with OmniFill's scrapbook, all but one participant eventually transitioned to a ``collect-then-fill'' strategy, in which multiple information demands were collected into the scrapbook without returning to the contact form in between.
This was true both of information that could be collected in a single selection (e.g. an entire address and phone number from one selection on a ``Contact Us'' page) and information from different pages, which was collected in multiple text selections often without returning to the contact form in between.
In most cases, participants did revisit the contact form before collecting \textit{all} of the demanded information, but after collecting multiple pieces of ``low-hanging fruit''.
Often, the last few straggling pieces of missing information were entered in smaller increments as the participants hunted them down.

Participants often lost track of which information they had collected already, and in some cases even expressed surprise when they saw how much information OmniFill had extracted -- ``wow, that already knocks out a lot.''

\subsubsection{Overcollection} Since it is OmniFill, and not the user, who needs to ``read through'' context in the scrapbook and extract relevant information, the main cost of adding new information to the bag of context occurs at the time of collection, and additional cost is largely not incurred even if the user's judgment of relevance turns out to be a false positive.
This distinguishes OmniFill from simple ``multiple copy-and-paste'' tools, which still require precision in foraging to prevent an overload of information to parse.
Participants frequently collected context with imprecise text selections (compared to the exact selection required for a copy-and-paste) or occasional duplicate information from different sources (e.g. collecting the phone number from the ``Contact Us'' page and from the website's footer).
Some participants suggested obviating manual collection entirely, but our own early prototype testing showed that OmniFill's success was significantly diminished when collecting \textit{all} browsing context indiscriminately, without manual foregrounding of information the user finds relevant.

By lowering the barriers for participants to ``overcollect'' information, the system was able in some cases to correct errors, even transparently.
For example, one participant collected the wrong phone number for the school (collecting instead the phone number of the school's Title IX coordinator), and OmniFill generated a suggestion with this phone number.
Before this participant returned back to the contact form, they found another page that contained additional information they were looking for, alongside the correct phone number for the school.
When the participant quickly selected the entire information block -- including the correct phone number -- OmniFill corrected its phone number suggestion, now that there were two phone numbers to choose from (and sufficient natural language information in its bag of context to determine which of the two was more appropriate), without the participant ever even seeing the incorrect suggestion.

Overcollection may also help when OmniFill makes an incorrect inference based on information not in the context, e.g. when incorrectly assuming a particular elementary school's served grade levels are K-5 or a high school's are 9-12.
One participant accepted one such incorrect suggestion and, upon later finding the school's enrollment total on a ``school profile'', collected the entire paragraph as context; since the school profile also included the true ``grade levels served'' field value ``K-6'', OmniFill updated its suggestion.
This was not a transparent correction, since the participant had already accepted the ``K-5'' suggestion, but after noticing the change, they updated the information in the contact form.

\subsection{Trust in OmniFill's suggestions}
Because OmniFill assumes the role of locating relevant information in its bag of context during the information-gathering task, users who want to collect context opportunistically and avoid keeping track of the bag of context must place trust in OmniFill's inferences.
We observed that users often trusted OmniFill's suggestions, even when that trust may have been unfounded.

The contact form contained a ``grade levels served'' field, which should be filled with e.g. ``9-12'' for a typical American high school.
When websites did not explicitly include this information, some participants were reluctant to make assumptions for the value of this field.
Although OmniFill often withheld suggestions unless the relevant information was explicitly present in its bag of context (for example, we never observed the street address field being incorrectly populated), the model did often return with a suggestion for the grade level field without the context including this information; this happened at least once for 15 of 18 participants.

It was rare, however, for participants to question this information when OmniFill suggested it; only one participant did explicitly reject this overeager grade level completion, and that participant had selected only a very small amount of text that they could immediately observe did \textit{not} contain the grade level information.
In a flipped example, one participant, who had filled out most of the contact form, was struggling to find the name of the school's principal.
They chose to select the name of an assistant principal for their scrapbook.
Only after heading back to the contact form and observing that OmniFill \textit{did not} offer a suggestion for the ``Principal'' field did the participant give up on finding the information and move onto the next school.

Although the data-formatting task contained fewer opportunities for participants to rely on OmniFill to make ``judgment calls'', we still observed two instances of participants accepting and saving OmniFill's incorrect suggestion.
For each of these two participants, OmniFill made the strange suggestion in one case to duplicate a correctly-formatted phone number into a blank phone number field.
Although both participants had already correctly handled this case in the past, they chose to accept this suggestion from OmniFill; one said aloud, ``Oh, yeah, sure. Why not?'' as they did this, indicating a willingness to allow OmniFill to take the reins in deciding and carrying out the task specification.

\subsection{Partial success}
OmniFill was not perfect, occasionally visibly failing to extract explicitly-foregrounded information in the information-gathering task or offering incorrect suggestions (or no suggestions at all) for the data-formatting task.

Still, this partial success offered a value-add in both tasks.
In the information-gathering task, participants could always fall back to traditional information-foraging techniques to find data that OmniFill failed to extract.
In the data-formatting task, OmniFill's success rate on offering correct T-Shirt sizes was high (across the 18 participants, OmniFill offered a correct suggestion in 306 of the 347 cases (= 88\%) where the T-Shirt size needed to be updated), so participants were often presented with at least one correct suggestion even when the system couldn't get every field right.

Even when the phone number suggestion was incorrect, we found that participants often \textit{accepted the incorrect suggestion} and then made changes to the field, rather than editing the original text field value.
Of the 16 participants who received a suggestion to change a phone number from one incorrect format into another incorrect format, 15 accepted one of these suggestions at least once (and 9 did this multiple times); Figure \ref{fig:task-2-success} indicates incorrect suggestions which were accepted by the user.
For example, a common failure mode for this task was for OmniFill to format the phone number correctly but without removing the country code (i.e. \texttt{+1 (\#\#\#) \#\#\#-\#\#\#\#}) or to hyphenate a ten-digit number (transforming \texttt{\#\#\#\#\#\#\#\#\#\#} to \texttt{\#\#\#-\#\#\#-\#\#\#\#}).
By accepting these suggestions, participants allowed OmniFill to get them \textit{closer} to their desired result for the form field.

Many participants noted that partial failure may not be harmless, however.
Ten participants, when asked about concerns using OmniFill in the real world, cited accuracy considerations.
Many noted, either during the interview or while completing the tasks, that although they may notice ``obvious'' errors or those that don't conform to the task specification, they might not notice if OmniFill inserted incorrect information in the information-gathering task or changed phone numbers in the data-formatting task.
Four participants, including some concerned about OmniFill's accuracy, still suggested that the potential for human error in a task like phone number formatting was high, and that they had more trust in OmniFill to be correct.

\subsection{Context visibility}

\subsubsection{Verifying suggestions}
LLMs, especially when asked to supply information not in their prompt context, can ``hallucinate'' incorrect information~\cite{dzirietal2022origin}.
Although OmniFill can in some cases offer value to users by suggesting general world or language knowledge not present in the model's context window (as with its school grade level suggestions, which were largely correct), we anticipate that, in many situations where OmniFill offers a value-add, the information present in the suggestion is present in the bag of context (either literally or in some pre-transformation state).
Since suggestion accuracy was often a concern for participants, offering better visibility into where information is coming from may help.

\subsubsection{Incorrect prior examples}
Because OmniFill observes much of the user's behavior passively and has a largely immutable bag of context, there is a risk that participants will accidentally teach the system incorrectly, saving an example that causes problems in the future (such as suggestions of incorrect information from prior examples or inferences of an incorrect task specification).
We did not provide users with a mechanism for deleting incorrectly-saved examples, and 9 of 18 participants saved at least one incorrect example during the data-formatting task; even those who corrected their errors were not able to delete the mistakenly-saved examples.

In two cases in the data-formatting task, we observed participants making a mistake that was later reflected in the model's output, both related to formatting phone numbers when two were present in the field.
Neither mistake was caused by the participant accepting a suggestion from OmniFill, but in both cases, when presented with a similar profile later in the task, OmniFill made the same error.

When we later called the GPT-4 API with the prompts from the data-formatting task, despite GPT-4's much higher general success rate on this task (depicted in Figure \ref{fig:task-2-success}), both of these later errors still manifested.
Although a more powerful LLM may be able to learn patterns and infer a task specification with few examples, this suggests that incorrect examples of prior user behavior can still be a critical source of ultimately incorrect predictions by the system.

%% file: 08-discussion.tex
LLM-backed systems like OmniFill can offer convenient form-filling suggestions in a single, domain-agnostic package without requiring significant task-specific configuration.
Because of this general applicability, 
we anticipate that, despite lower-precision task specifications and the risk of difficult-to-notice errors, this type of predictive system will become increasingly integrated into daily computing, just as search result ranking and mobile predictive keyboards have.

System designers must understand both the limitations of LLM-backed approaches and the character of users' interactions with such systems.
In this section, we discuss implications for privacy and accuracy, especially through the lens of potential future directions of this work.

\subsection{Privacy and security}
When explicitly asked about real-world concerns about using OmniFill, only six of 18 participants mentioned privacy or information security as a consideration, even though the data-formatting task involved working with simulated personal information.
We did not explain to participants upfront that OmniFill does not run entirely on the user's local computer, but even users who know they are interacting with an online system may not consistently protect the privacy of their information \cite{sundar2013unlocking}.
Systems that allow users to ingest \textit{other} peoples' personal information must be considered even more strictly.
A system like OmniFill benefits from passively observing the user's behavior or collecting implicit context (e.g. bringing in text immediately before and after a user's text selection in case their mouse cursor ``aim'' was imperfect), but this requires trust in the remote LLM API used by the system.
Because OmniFill is designed to interoperate between siloed ecosystems, privacy concerns persist even outside of systems that are known to track user behavior.

Running general suggestion systems locally on the user's computer may become feasible as model architectures grow more efficient and computers become more powerful, keeping model queries private.
Still, future work may choose to store many prior user examples (more than what can fit into a single LLM prompt) for later retrieval during prompt construction (as in~\cite{levy2023diverse}).
System designers should be careful when collecting and opaquely storing browsing context long-term, even locally, since this practice can increase the consequences of a system breach.

\subsection{Accuracy}
Even after the system appears to have ``learned'' a task, the lack of a rigorous task specification can cause occasional errors (as Figure \ref{fig:task-2-success} demonstrates in the GPT-4 section).
OmniFill is designed to assist users in their manual tasks, not construct reliable automations.
However, ``automation bias'' is known to result in uncertainty among users of automated systems~\cite{BOND2018S6}, and a ``good-enough'' automation of menial tasks may cause the user to ``check out'' and stop manually verifying the results of suggestions, as reported by some participants after completing our data-formatting task.
As Figure \ref{fig:task-2-success} illustrates, success can be highly dependent on choice of model, depending on the task at hand.
This suggests challenges for developers building systems that are not entirely under their control.

Since users cannot inspect the LLM's workings, it is difficult to form a mental model of which \textit{types} of errors the LLM is likely to make, meaning even partial success of the system may have variable utility.
We observed users accepting OmniFill's ``judgment calls'' even though the system had never been told how it should make these judgments; the ``authority'' of the system seemed persuasive even in those very situations where an automated system cannot know how to behave.

Future improvements to such systems may further reduce the cognitive load demanded of users during certain tasks (e.g. by improving implicit context collection so that less information needs to be explicitly foregrounded), but these exact improvements may cause users to be less likely to notice when the model makes an error.
Our results from the information-gathering task demonstrate that the system is already powerful enough to permit users to collect and use information without noticing that they had collected it.
How, then, would a user in this situation notice if the suggestion is incorrect?

Context visibility may play a key role.
Because the ``collect-then-fill'' strategy recruits OmniFill's bag of context as an auxiliary ``memory'' for the user, future work in constructing interface features could allow users to \textit{peer into} the bag of context and view context sources \textit{in situ}.
Since information used in suggestions is often present in the bag of context, either in some literal or pre-transformation form, these features could surface just the context that matches with OmniFill's suggestions.
Although a literal search for each suggestion in the bag of context may not yield results when the task involves transformation operations, additional prompts to a lighter-weight LLM (or a semantic search using an embedding model) could be used to handle many simpler cases.
OmniFill's full LLM prompt would still be responsible for producing the actual suggestions, since the full bag of context may be valuable for making high-quality predictions (e.g. if the form structure is poorly-labeled but can be learned through prior form-filling examples), but post hoc ``attribution'' may be achieved through an auxiliary system.

Although \textit{in situ} training of the system can be a low-friction way to offer predictive suggestions, systems should provide users with the ability to view and refine their task specifications, e.g. by curating their set of examples to maximize system accuracy.
Future work could assist users in this process by detecting and surfacing potentially anomalous prior examples or by engaging the user in a dialogue to define and fine-tune task specifications as the system is used over time.

%% file: 09-conclusion.tex
Not every task calls for full automation or an elaborate specification.
Even when task definitions are fuzzy, partial automation of the simpler tedious components of form filling tasks can prove valuable, and LLM-backed systems like OmniFill can serve as a ``glue'' between arbitrary context sources and target forms without heavy configuration.
We demonstrate opportunities of LLM-backed systems to assist in a unique subspace of form filling tasks, then describe our observations of users trying the prototype.
We believe this is a rich space for future system designers to explore, but care must be taken to understand how people perceive and use such systems, especially in a landscape of rapidly-expanding capabilities and expectations for artificial intelligence tools.